\documentclass[12pt,a4paper]{article}
\usepackage[utf8]{inputenc}
\usepackage[T1]{fontenc}
\IfFileExists{lmodern.sty}{\usepackage{lmodern}}{}
\usepackage[british]{babel}
\usepackage{amsmath,amssymb,amsthm,mathtools}
\usepackage{graphicx}
\usepackage{booktabs,tabularx,multirow,longtable,array}
\usepackage{caption}
\usepackage{float,placeins}
\usepackage{enumitem}
\usepackage{setspace}
\usepackage[round,authoryear]{natbib}
\usepackage[margin=2.5cm]{geometry}
\usepackage[hidelinks]{hyperref}
\theoremstyle{definition}

\title{\textbf{Stablecoin Runs as Safe-Asset Stress Signals: Episode-Level Evidence for Central Bank Surveillance}}
\author{Ahmad Qadi\thanks{Corresponding author. Email: \texttt{ahmad.qadi.22@ucl.ac.uk}.}\quad Hui Gong\quad Francesca Medda\\[0.6em]
\normalsize UCL Institute of Finance and Technology\\ \normalsize University College London, London, United Kingdom}
\date{}

\begin{document}
\maketitle
\thispagestyle{empty}

\begin{abstract}
\noindent This paper asks whether stablecoin runs carry information about U.S.\ Treasury yields. Using daily data on USDT, USDC and BUSD from 2021 to 2025 and five stress episodes, it measures flows from changes in token units rather than market capitalisation, which removes the phantom inflows the standard measure records when tokens lose their peg. Flows have no stable relationship with yields in normal periods. During the three panic-driven runs, a one-billion-dollar five-day redemption is associated with an 8.5 basis point fall in the three-month bill yield within ten trading days, while two regulatory episodes carry no such signal. The response is broad across maturities and has the opposite sign to what reserve liquidation predicts, which points to flight to quality. The effect is led by the TerraUSD collapse and corroborated by the USDC/SVB run. Five-day redemptions of roughly \$2 to 3 billion during a panic mark the scale at which stablecoin stress becomes informative for surveillance.
\end{abstract}



\newpage
\section{Introduction}\label{c3:sec:intro}

Fiat-backed dollar stablecoins grew from under USD~20 billion in early 2020 to over USD~160 billion by early 2025. The largest issuers, Tether (USDT) and Circle (USDC), hold their reserves mainly in U.S.\ Treasury bills, reverse repurchase agreements and other liquid dollar assets. Stablecoins therefore sit on the boundary between digital-asset markets and the sovereign safe-asset market: they are issued and redeemed inside the crypto ecosystem, but their backing is drawn from the instruments that matter for monetary policy implementation, collateral markets and central bank surveillance.

That boundary matters most under stress. In normal times stablecoin issuance and redemption are a small source of marginal demand for safe assets. During a run, redemptions become large, fast and concentrated, and the natural concern is that issuers sell reserves and move Treasury prices. But stablecoin runs rarely happen alone. They coincide with wider risk aversion, crypto-market failures, banking stress and flights to quality. The question for a central bank is therefore not only whether redemptions mechanically push yields up, but whether stablecoin stress carries usable information about the state of the safe-asset market, and if so, when.

The closest existing evidence is \citet{AhmedAldasoro2025}, who estimate that stablecoin inflows lower short-term Treasury yields on average across 2021 to 2025 and interpret the result as a reserve-flow channel. This paper starts from the same question and changes the focus from the average day to the stress regime. For surveillance, what matters is how the system behaves when confidence breaks and the stablecoin sector becomes information-sensitive, not the average relationship across calm days.

\begin{figure}[H]
\centering
\includegraphics[width=\textwidth]{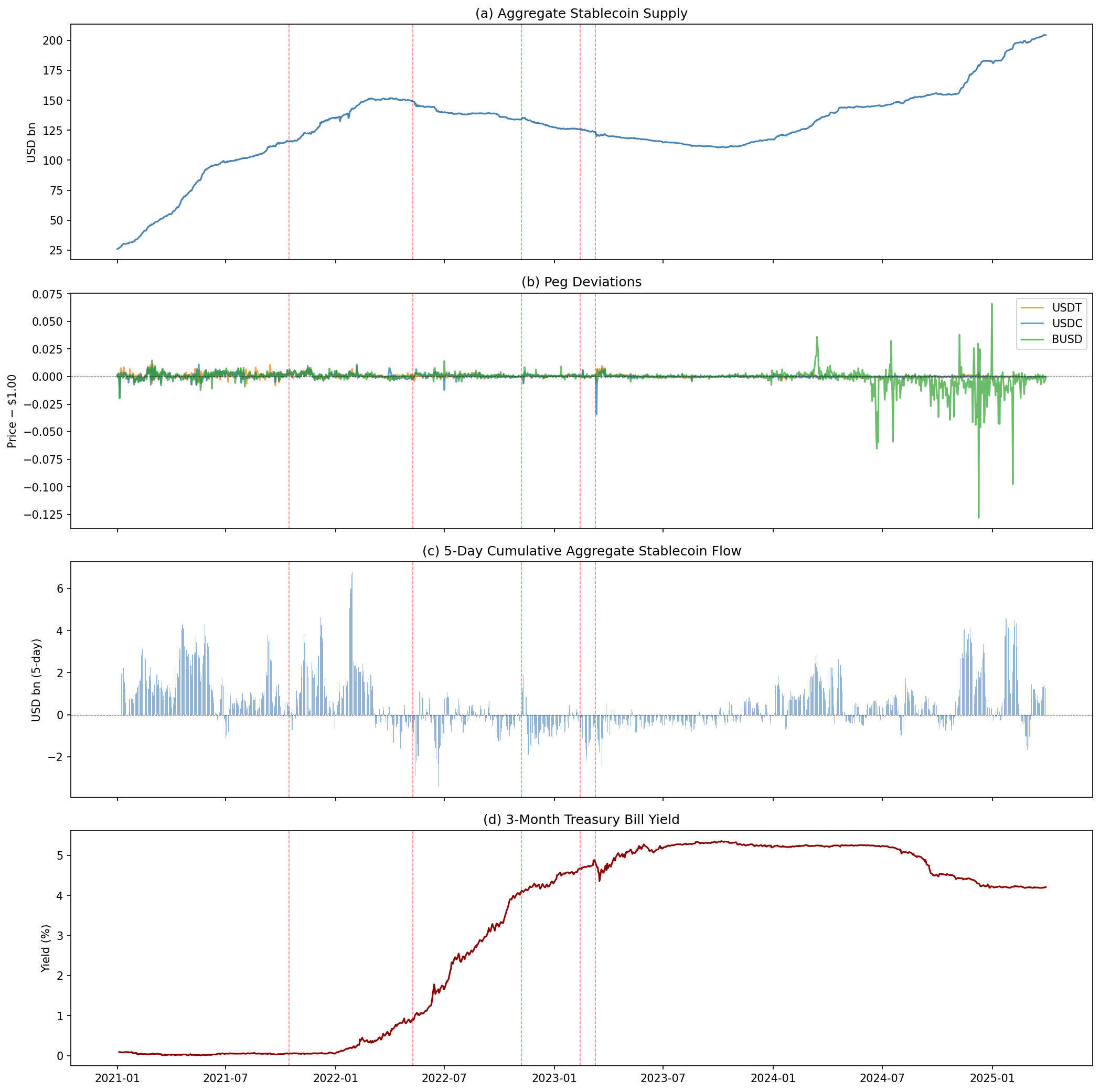}
\caption{Overview, 2021--2025. (a) Aggregate supply of USDT, USDC and BUSD. (b) Daily peg deviations. (c) Five-day cumulative aggregate flow, price-corrected. (d) Three-month Treasury bill yield. Vertical lines mark the five stress episodes.}
\label{c3:fig:overview}
\end{figure}

Figure~\ref{c3:fig:overview} shows the setting. Five episodes between late 2021 and spring 2023 combine large changes in stablecoin supply with visible peg stress and clear narrative timing, and they are of two kinds: three panic-driven runs and two regulatory actions. That distinction turns out to be the key to the results.

The paper makes four contributions.

First, a measurement correction. Existing work measures stablecoin flows from changes in market capitalisation. For a token that has lost its peg, market capitalisation can rise because the price recovers while tokens are still being redeemed, so the standard measure records phantom inflows at exactly the moments that matter. Flows are measured here from changes in token units outstanding at par, which recovers the dollar amount issuers actually have to pay out. Around the USDC/SVB episode the correction changes the measured direction of flows.

Second, state-dependent evidence. A two-state Markov-switching model finds a distinct stress regime, and local projections with run interactions show that the flow-yield relationship is absent in normal periods and economically meaningful during panic-driven runs.

Third, a finding about episode type. Splitting the five episodes into panic-driven and regulatory, a distinction fixed before the results were known, shows that the signal comes entirely from panic. Regulatory wind-downs, however large, leave no reliable trace in yields. This is what a surveillance framework needs to know: not every large redemption is informative, only those driven by a loss of confidence.

Fourth, a reading of the mechanism that supports a surveillance rather than a transmission interpretation. The sign of the effect is the opposite of what reserve liquidation would produce, and the response is spread across the curve rather than concentrated in bills. Panic-driven stablecoin redemptions therefore behave as a high-frequency marker of a broader flight to quality, available daily and before slower macro-financial indicators confirm that a stress state has begun. Five-day redemptions of roughly \$2 to 3 billion are the scale at which that marker becomes informative.

Section~\ref{c3:sec:lit} reviews the literature. Section~\ref{c3:sec:data} describes the data, the flow correction and the episodes. Section~\ref{c3:sec:method} sets out the empirical strategy and the logic behind each step. Section~\ref{c3:sec:results} presents and discusses the results, Section~\ref{c3:sec:policy} draws out the implications for central bank surveillance, and Section~\ref{c3:sec:conclusion} concludes.

\section{Literature Review}\label{c3:sec:lit}

Four literatures bear on the question: safe-asset demand and sovereign yields, run risk in money-like instruments, the empirics of stablecoin markets, and the policy debate on stablecoin oversight. Each supplies one piece of the design below.

Demand for safe, liquid assets moves sovereign yields beyond what expectations and term premia explain. \citet{KrishnamurthyVissing2012} show that the supply of Treasury debt is inversely related to the spread between Treasuries and other high-grade bonds, which they read as the convenience yield investors pay for safety and liquidity; \citet{KrishnamurthyVissing2015} find the same channel at work in Federal Reserve purchases, and \citet{CaballeroFarhi2018} place it in a global safe-asset shortage in which institutions that hold safe assets as backing for money-like claims affect equilibrium rates. The maturity dimension matters for what follows. In the preferred-habitat model of \citet{Greenwood2015}, yields depend on the maturity composition of supply relative to demand, so an issuer selling Treasury bills to meet redemptions should move bill yields most. That is a testable prediction, and the maturity and spread tests below are built to test it. \citet{Sunderam2015} brings the argument closest to stablecoins: when short-term money-like claims expand, intermediaries absorb safe collateral, and when they contract, collateral is released. Stablecoin issuance and redemption map directly onto that framework. \citet{Nagel2016} shows that the convenience yield on near-money assets depends on the level of short rates, \citet{JiangKrishnamurthy2021} document that Treasury convenience yields move in stress, and \citet{HeKrishnamurthy2013} show that intermediary balance-sheet constraints amplify price responses to demand shocks. Together these papers say that the relationship, if it exists, should be state-dependent, which is why the design below does not estimate a single average effect.

The second literature explains why stablecoins should run at all. Since \citet{DiamondDybvig1983}, any claim that promises par redemption against assets whose liquidation value can become uncertain is exposed to self-fulfilling runs. \citet{Covitz2013} document runs on asset-backed commercial paper conduits and \citet{GortonMetrick2012} read rising repo haircuts in 2008 as a run on the shadow banking system. Money market funds are the closest analogy: \citet{SchmidtTimmermann2016} show that the 2008 run on prime funds was accompanied by flows into government-only funds, a flight to quality within money, which is the pattern seen when USDC redemptions in March 2023 coincided with inflows to USDT. \citet{DangGortonHolmstrom2017} explain the trigger: money-like claims are valuable because they are information-insensitive, and a run begins when a claim becomes information-sensitive, whether through an exchange failure, a bank exposure, a regulatory action or a visible peg break. \citet{AhmedAldasoroDuley2024} formalise this for stablecoins in a global-games setting, and \citet{MaZengZhang2025} document the two-tier redemption structure that shapes stablecoin runs in practice: only a few institutional participants redeem directly with the issuer, while everyone else trades in secondary markets, so peg stability depends on arbitrageurs as much as on reserves. Stablecoins settle continuously, with no bank holiday, deposit insurance or lender of last resort, which is why the design below works at daily frequency over short horizons.

The stablecoin empirics are recent and thin. \citet{AhmedAldasoro2025} is the direct predecessor: instrumented local projections on aggregate dollar-backed supply show that inflows lower short-term yields and outflows raise them more strongly, concentrated at short maturities and read as reserve flows. This paper departs from it in three ways: it estimates state-dependent rather than average effects, it replaces market-capitalisation flows with a units-based measure, and it exploits the episodes directly through run interactions, episode decompositions and placebo windows. \citet{LyonsViswanath2023} show that reserve quality, arbitrage responsiveness and issuer reputation govern peg recovery; \citet{GortonZhang2023} draw the historical parallel with wildcat banking and argue for treating stablecoins as privately issued money; \citet{Catalini2022} set out the design choices, from reserve composition to the holder's legal claim on the issuer, that determine how a token behaves under redemption pressure. \citet{AlAsadi2025} study how stablecoins respond to macro-financial shocks, which is the reverse direction from the one studied here. \citet{Auer2025} and \citet{Reuter2025} document the growing cross-border role of stablecoin flows. None of this work estimates the episode-level relationship between redemptions and sovereign yields.

The policy literature treats stablecoins as a financial-stability matter but stops short of calibration. The BIS argues that large foreign-currency stablecoins threaten monetary sovereignty and invite regulatory arbitrage \citep{BIS2022AER, BIS2023}; the Financial Stability Board recommends full reserve backing, transparent disclosure and reliable par redemption \citep{FSB2023}; the IMF places crypto-asset oversight within financial-stability policy \citep{IMF2021}. What none of them says is when stablecoin flows become large enough to matter for a central bank watching the Treasury market. That is the gap this paper addresses: it tests whether the relationship is meaningful specifically in runs, corrects the flow measure where it fails, and reports the scale of redemptions at which stablecoin stress becomes informative.

\section{Data and Variable Construction}\label{c3:sec:data}

The sample is a daily business-day panel from 4~January 2021 to 31~March 2025, covering the growth of the stablecoin market, the 2022 to 2023 tightening cycle and the five episodes. Daily price, market capitalisation and volume for USDT, USDC and BUSD come from CoinGecko. Daily yields on three-month bills (FRED DTB3), six-month bills (DTB6), two-year notes (DGS2) and ten-year notes (DGS10) come from FRED; the three-month bill is the main dependent variable and the others are used for the maturity and spread tests. Controls are the effective federal funds rate (DFF), the VIX (VIXCLS) and the daily S\&P~500 return. All series are aligned to Treasury trading days.

\paragraph{Flows and the price correction.} The standard proxy for stablecoin issuance and redemption is the change in market capitalisation. For a token with price $P_{i,t}$ and $N_{i,t}$ units outstanding, $\text{MCap}_{i,t} = P_{i,t} N_{i,t}$, so
\begin{equation}\label{c3:eq:mcap_decomp}
    \Delta \text{MCap}_{i,t} \;=\; P_{i,t} \cdot \Delta N_{i,t} \;+\; N_{i,t-1} \cdot \Delta P_{i,t}.
\end{equation}
The first term is the change in units, which is what issuance and redemption are. The second is a pure price effect. When the token trades at par it is negligible, and the standard proxy is fine. When the token loses its peg it is not: market capitalisation can rise because the price recovers while units are still being redeemed. Flows are therefore measured from units outstanding, valued at par:
\begin{equation}\label{c3:eq:corrected_flow}
    \Delta N_{i,t} \;=\; \frac{\text{MCap}_{i,t}}{P_{i,t}} - \frac{\text{MCap}_{i,t-1}}{P_{i,t-1}},
\end{equation}
in billions of dollars. Table~\ref{c3:tab:svb_compare} shows why this matters, using the days around the USDC de-peg. On 13~March 2023 the market-capitalisation measure records an inflow of \$0.88 billion; the units measure shows that redemptions continued that day at \$0.27 billion, and the apparent inflow is nothing but the price recovering from \$0.966 to \$0.994. Over the full sample the two measures have almost identical means and standard deviations, and differ only in the tails: the corrected five-day aggregate flow reaches $-\$3.41$ billion against $-\$2.91$ billion for the raw measure. The error therefore sits exactly in the windows a study of runs cares about. All flow regressions use the corrected measure.

\begin{table}[H]
\centering
\small
\caption{USDC flow around the SVB de-peg: market-capitalisation versus units-based measure.}\label{c3:tab:svb_compare}
\begin{tabular}{lcrr}
\toprule
\textbf{Date} & \textbf{USDC price} & \textbf{Raw $\Delta$MCap (\$bn)} & \textbf{Corrected $\Delta N$ (\$bn)} \\
\midrule
8 Mar 2023  & 1.0006 & $-$0.18 & $-$0.19 \\
9 Mar 2023  & 1.0008 & $-$0.05 & $-$0.06 \\
10 Mar 2023 & 1.0015 & $-$0.07 & $-$0.10 \\
11 Mar 2023 & 0.9935 & $-$1.24 & $-$0.90 \\
12 Mar 2023 & 0.9656 & $-$2.78 & $-$1.65 \\
13 Mar 2023 & 0.9937 & \textbf{+0.88} & $-$0.27 \\
14 Mar 2023 & 0.9994 & $-$0.78 & $-$1.02 \\
15 Mar 2023 & 1.0008 & $-$1.09 & $-$1.14 \\
\bottomrule
\end{tabular}
\smallskip

\noindent\footnotesize{The raw measure records an inflow on 13~March because USDC's price recovered; the units-based measure shows continued redemption.}
\end{table}

\paragraph{Flow variables.} The daily aggregate flow is $\text{AggFlow}_t = \sum_i \Delta N_{i,t}$, positive for net issuance and negative for net redemption, and the five-day cumulative flow is $\text{AggFlow}^{5d}_t = \sum_{k=0}^{4} \text{AggFlow}_{t-k}$. Because a signed flow makes the coefficient on outflows awkward to read, the main regressions use redemption magnitudes: $\text{Redemption}_t = \max(-\text{AggFlow}_t,0)$ and $\text{Redemption}^{5d}_t = \max(-\text{AggFlow}^{5d}_t,0)$, so that a value of one is a one-billion-dollar net redemption and a negative coefficient means redemptions go with lower yields. The five-day measure is the headline regressor because it captures sustained redemption pressure rather than one day's noise. Peg deviation is $P_{i,t} - 1$. The cumulative yield change at horizon $h$ is $\Delta_h y_{t+h} = y_{t+h} - y_{t-1}$. Flow and peg variables are winsorised at the 1st and 99th percentiles; results are unchanged without winsorising. Table~\ref{c3:tab:variables} collects the definitions.

\begin{table}[H]
\centering
\small
\caption{Variable definitions.}\label{c3:tab:variables}
\begin{tabularx}{\textwidth}{llX}
\toprule
\textbf{Variable} & \textbf{Notation} & \textbf{Definition} \\
\midrule
Peg deviation & $\text{PegDev}_{i,t}$ & $P_{i,t} - 1.00$ (USD) \\
Daily flow (units) & $\Delta N_{i,t}$ & $\text{MCap}_{i,t}/P_{i,t} - \text{MCap}_{i,t-1}/P_{i,t-1}$ (\$bn) \\
Aggregate flow & $\text{AggFlow}_t$ & $\sum_i \Delta N_{i,t}$ (\$bn); positive is net issuance \\
5-day cumulative flow & $\text{AggFlow}^{5d}_t$ & $\sum_{k=0}^{4} \text{AggFlow}_{t-k}$ (\$bn) \\
Redemption magnitude & $\text{Redemption}_t$ & $\max(-\text{AggFlow}_t,0)$ (\$bn) \\
5-day redemption magnitude & $\text{Redemption}^{5d}_t$ & $\max(-\text{AggFlow}^{5d}_t,0)$ (\$bn) \\
Cumulative yield change & $\Delta_h y_{t+h}$ & $y_{t+h} - y_{t-1}$ (percentage points) \\
Run dummy & $\text{Run}_t$ & 1 inside any of the five event windows \\
\bottomrule
\end{tabularx}
\end{table}

\paragraph{The five episodes.} Table~\ref{c3:tab:events} defines the episodes. In October 2021 the CFTC settled with Tether and Bitfinex, fining them \$41 million and \$1.5 million for misleading statements about USDT's backing: a regulatory strike over reserves, not a market panic. In May 2022 TerraUSD collapsed and the contagion reached fiat-backed tokens, with USDT briefly below par and large net redemptions. In November 2022 FTX failed and confidence in crypto-market infrastructure broke across several tokens at once. In February 2023 the New York Department of Financial Services ordered Paxos to stop minting BUSD, beginning an orderly wind-down. In March 2023 Circle disclosed that \$3.3 billion of USDC reserves sat at Silicon Valley Bank, USDC traded down to \$0.87, and a wave of redemptions followed. Three of the five are panic-driven runs (Terra, FTX, USDC/SVB) and two are regulatory actions (CFTC/Tether, BUSD). Each window runs from the event's start to the point at which the narrative record has the episode abating: 41 or 42 business days for the four 2022 to 2023 episodes and 29 for the CFTC/Tether settlement, whose market reaction was shorter. The windows were fixed from the narrative before any regression was run and were not tuned against the results. The combined run dummy covers 171 business days, about 16\% of the regression sample. The BUSD and SVB windows overlap for about five weeks; the CFTC/Tether window stands alone in late 2021.

\begin{table}[H]
\centering
\small
\caption{Stress episode definitions.}\label{c3:tab:events}
\renewcommand{\arraystretch}{1.2}
\begin{tabularx}{\textwidth}{>{\raggedright\arraybackslash}p{2.7cm} p{1.9cm} >{\raggedright\arraybackslash}p{3.7cm} c >{\raggedright\arraybackslash}X}
\toprule
\textbf{Episode} & \textbf{Type} & \textbf{Window} & \textbf{Days} & \textbf{Shock} \\
\midrule
CFTC/Tether settlement & Regulatory & 15 Oct -- 26 Nov 2021 & 29 & CFTC fines Tether \$41m and Bitfinex \$1.5m over reserve misrepresentation \\
TerraUSD collapse & Panic & 9 May -- 8 Jul 2022 & 42 & Algorithmic stablecoin failure; contagion to USDT \\
FTX collapse & Panic & 7 Nov 2022 -- 6 Jan 2023 & 41 & Exchange insolvency; crypto confidence shock \\
BUSD enforcement & Regulatory & 13 Feb -- 14 Apr 2023 & 42 & NYDFS orders Paxos to stop minting BUSD \\
USDC/SVB de-peg & Panic & 10 Mar -- 9 May 2023 & 41 & Circle discloses \$3.3bn at SVB; USDC trades at \$0.87 \\
\bottomrule
\end{tabularx}
\end{table}

\paragraph{Summary statistics.} Table~\ref{c3:tab:sumstats} describes the sample. USDT's largest absolute peg deviation is about one cent; BUSD's reaches 12.8 cents during its wind-down. Aggregate daily flow has a standard deviation of about \$400 million with extremes beyond \$2.5 billion in both directions, and five-day redemptions reach \$3.4 billion. Those extremes are the episodes.

\begin{table}[H]
\centering
\small
\caption{Summary statistics, business days, 2021--2025.}\label{c3:tab:sumstats}
\begin{tabular}{lrrrrr}
\toprule
& \textbf{Mean} & \textbf{Std} & \textbf{Min} & \textbf{Max} & \textbf{N} \\
\midrule
\multicolumn{6}{l}{\emph{Stablecoin variables (price-corrected flows)}} \\
PegDev USDT & 0.0005 & 0.0018 & $-$0.0081 & 0.0114 & 1{,}061 \\
PegDev USDC & 0.0004 & 0.0019 & $-$0.0191 & 0.0110 & 1{,}061 \\
PegDev BUSD & $-$0.0009 & 0.0089 & $-$0.1281 & 0.0662 & 1{,}061 \\
$\Delta N$ USDT (\$bn) & 0.080 & 0.268 & $-$3.432 & 1.472 & 1{,}061 \\
$\Delta N$ USDC (\$bn) & 0.024 & 0.289 & $-$1.142 & 4.494 & 1{,}061 \\
$\Delta N$ BUSD (\$bn) & $-$0.003 & 0.155 & $-$3.187 & 0.554 & 1{,}061 \\
AggFlow (\$bn/day) & 0.101 & 0.404 & $-$2.535 & 4.334 & 1{,}061 \\
AggFlow 5-day (\$bn) & 0.500 & 1.247 & $-$3.414 & 6.765 & 1{,}057 \\
Redemption 5-day (\$bn) & 0.213 & 0.393 & 0.000 & 3.414 & 1{,}057 \\
\midrule
\multicolumn{6}{l}{\emph{Yields and controls}} \\
3M T-bill (\%) & 3.094 & 2.202 & 0.010 & 5.360 & 1{,}061 \\
6M T-bill (\%) & 3.152 & 2.123 & 0.020 & 5.360 & 1{,}061 \\
2Y Treasury (\%) & 3.112 & 1.799 & 0.090 & 5.190 & 1{,}061 \\
10Y Treasury (\%) & 3.215 & 1.181 & 0.930 & 4.980 & 1{,}061 \\
VIX & 19.38 & 5.18 & 11.86 & 38.57 & 1{,}059 \\
S\&P 500 return & 0.0004 & 0.0104 & $-$0.043 & 0.055 & 1{,}058 \\
\bottomrule
\end{tabular}
\end{table}

\section{Methodology}\label{c3:sec:method}

\subsection{What the Design Has to Distinguish}

Three channels could connect stablecoin redemptions to Treasury yields, and the design is built around what each one predicts.

The first is direct reserve selling: issuers meet redemptions by selling bills, and the selling moves bill prices. This is the mechanism \citet{AhmedAldasoro2025} read into their estimates. It has a specific signature: yields should rise, the effect should sit at the short maturities where reserves are held, and it should scale with the volume sold. Scale is already a problem for it. Five-day redemptions in this sample never exceed \$3.4 billion, while marketable Treasury debt grew from about \$22 trillion to \$28 trillion over the period and daily Treasury trading volume exceeded \$900 billion by 2024 \citep{SIFMA2025}. A few billion of reserve sales is small against that depth, so a price impact of the size reported below is hard to attribute to liquidation volume alone.

The second is a common shock: a crypto-market failure, a bank exposure or a general flight to safety drives both the run and the yield move, and neither causes the other. The third is a surveillance signal: redemptions do not cause yield moves and may not add information beyond other risk indicators, but they are fast, granular and observable every day, so they mark the onset of a stress state before slower indicators confirm it.

The second and third channels make the same prediction, that redemptions and yield moves co-move in stress, and no test here separates them; the paper does not claim to. The first channel is different, because its signature can be tested: the sign of the effect, its location on the curve, and whether it is confined to the bills reserves actually hold. That is what the maturity and spread tests are for. The surveillance interpretation does not need the second and third channels to be separated, since a central bank benefits from a fast indicator of a stress state whether or not the indicator is itself a cause.

\subsection{Sequence of Tests}

The empirical strategy follows a sequence in which each step answers a question raised by the one before. Rolling regressions ask whether the flow-yield relationship varies over time at all. A Markov-switching model asks whether that variation takes the form of a distinct stress regime. Local projections with run interactions then estimate the dynamic response in normal and run periods and are the main estimator. Maturity and spread tests ask which channel the response is consistent with. The episode-type split, episode decomposition and leave-one-out checks ask whether the pooled estimate hides heterogeneity and how much of it rests on one episode. Event-window, placebo-window and calendar-exclusion tests ask whether the window definition or coincident announcements are doing the work. The USDC/SVB comparison adds issuer-level evidence. Every method is standard in macro-finance; what is specific to the paper is the role each is given and the order in which they are read.\footnote{An instrumental-variables version following \citet{AhmedAldasoro2025} was also estimated. Instrumenting five-day flow with the five-day Bitcoin return and its volatility gives a first-stage $F$ of 29; the resulting full-sample estimate is $-0.0002$ ($p = 0.96$), which matches the absence of an average relationship reported in Section~\ref{c3:sec:results}. The instrument cannot be applied by regime, since crypto-market returns move for reasons unrelated to the runs that identify the results here, so the IV is not reported further.}

\subsection{Rolling Regressions and Break Detection}

The first step estimates
\begin{equation}\label{c3:eq:rolling}
    \Delta y_t = \alpha + \beta \cdot \text{AggFlow}_t^{5d} + \boldsymbol{\gamma}' \mathbf{Z}_t + \varepsilon_t,
\end{equation}
over 120-day rolling windows, where $\Delta y_t$ is the daily change in the three-month bill yield and $\mathbf{Z}_t = (\Delta \text{VIX}_t, \text{SP500ret}_t, \Delta \text{FFR}_t)'$. Binary segmentation \citep{BaiPerron1998} is applied to the rolling $\hat{\beta}$ sequence to locate breaks. The purpose is descriptive: if the coefficient is stable, there is no case for a state-dependent model; if it moves around the episodes, there is.

\subsection{Markov-Switching Regression}

Time variation could be gradual drift rather than a switch between states. A two-state Markov-switching regression \citep{Hamilton1989},
\begin{equation}\label{c3:eq:msw}
    \Delta y_t = \mu(S_t) + \beta(S_t) \cdot z_{\text{Flow},t} + \boldsymbol{\gamma}(S_t)' \mathbf{z}_t + \varepsilon_t,
    \quad
    \varepsilon_t \sim N(0, \sigma^2(S_t)),
\end{equation}
with latent state $S_t \in \{0,1\}$ and transition probabilities $p_{ij}$, tests for a distinct high-volatility regime; regressors are standardised, and the stress state is identified after estimation as the one with the larger residual variance. The model establishes whether a stress state exists and when it is active. It is not used to estimate the flow effect, because a two-regime model with a handful of regressors cannot trace a dynamic response over horizons.

\subsection{State-Dependent Local Projections}

The main estimates come from local projections \citep{Jorda2005}, which trace the response of yields to a redemption at each horizon without imposing a dynamic model, and which extend naturally to state dependence \citep{RameyZubairy2018}. For $h = 0, \ldots, 20$ business days,
\begin{equation}\label{c3:eq:lp}
    \Delta_h y_{t+h}
    =
    \alpha_h
    + \beta_h X_t
    + \theta_h \text{Run}_t
    + \delta_h (X_t \times \text{Run}_t)
    + \boldsymbol{\gamma}_h' \mathbf{Z}_t
    + \sum_{l=1}^{5} \phi_{h,l} \Delta y_{t-l}
    + \varepsilon_{t+h},
\end{equation}
where $X_t = \text{Redemption}^{5d}_t$ in the headline specification, $\text{Run}_t$ is the run dummy, $\beta_h$ is the normal-period response, $\beta_h + \delta_h$ the run-period response and $\delta_h$ the test of whether the two differ. Controls are $\Delta$VIX, the S\&P~500 return, $\Delta$FFR and five lags of the yield change; standard errors are Newey--West with bandwidth $h+1$. Two quantities are reported throughout: $\beta_{\text{run}} = \beta_h + \delta_h$, the total effect of a one-billion-dollar five-day redemption during a run, and $\delta_h$, the difference from normal periods. A positive $\beta_{\text{run}}$ would support reserve liquidation; a negative one would not.

\subsection{Maturity Segmentation and Spread Tests}

The headline projection is re-estimated with the six-month, two-year and ten-year yields as dependent variables, and then with the spreads $y^{6M} - y^{3M}$, $y^{2Y} - y^{3M}$ and $y^{10Y} - y^{3M}$. If reserve sales drove the result, the response should be concentrated in bills and the spreads should show the short end moving relative to the rest of the curve. If the response is broad and the spreads show no distinct short-end movement, the evidence favours flight to quality.

\subsection{Episode Type, Episode Decomposition and Leave-One-Out}

The five episodes are of two kinds, and pooling them assumes they carry the same signal. The headline specification is therefore re-estimated with the run dummy restricted to the three panic episodes and, separately, to the two regulatory episodes. The split was fixed when the CFTC/Tether episode was added to the sample, before the subsample results were known, so it is a pre-specified heterogeneity check rather than a search. Episode-by-episode projections then replace the pooled interaction with one interaction per episode, and a leave-one-out check drops each episode from the run dummy in turn. Both are needed because Terra is by far the largest run in the sample and could carry the pooled result on its own.

\subsection{Event Windows, Placebo Windows and Calendar Exclusions}

Three checks address the window definition. Forward-looking windows of 5, 10, 20 and 42 business days from each event start date, a construction separate from the narrative windows of Table~\ref{c3:tab:events}, show how the result builds as more of the stress period is included. Matched placebo windows draw random non-event periods with the same number of run observations as each specification and compare the actual run coefficient with the placebo distribution; both the lower-tail and the two-sided placebo $p$-value are reported and read together. Calendar exclusions re-estimate the headline projection without FOMC decision days and CPI release days, without the acute banking-stress window of 8 to 24~March 2023, and without both, to rule out the possibility that the run estimate is picking up announcement days or the SVB bank panic itself.

\subsection{Issuer-Level Comparison Around SVB}

USDC held reserves at Silicon Valley Bank and USDT did not, which gives an issuer-specific comparison:
\begin{equation}\label{c3:eq:svb}
    \Delta y_t
    =
    \alpha
    + \beta_1 \text{Post}_t
    + \beta_2 \Delta N_{\text{USDC},t}
    + \beta_3 (\Delta N_{\text{USDC},t} \times \text{Post}_t)
    + \boldsymbol{\gamma}' \mathbf{Z}_t
    + \sum_{l=1}^{5} \phi_l \Delta y_{t-l}
    + \varepsilon_t,
\end{equation}
with $\text{Post}_t = 1$ from 10~March 2023, estimated with daily and five-day USDC flows and with USDT as a placebo. Because the same weeks contain broad banking stress and large Treasury moves, this is read as supporting evidence rather than as a difference-in-differences design.

\section{Results and Discussion}\label{c3:sec:results}

\begin{figure}[H]
\centering
\includegraphics[width=\textwidth]{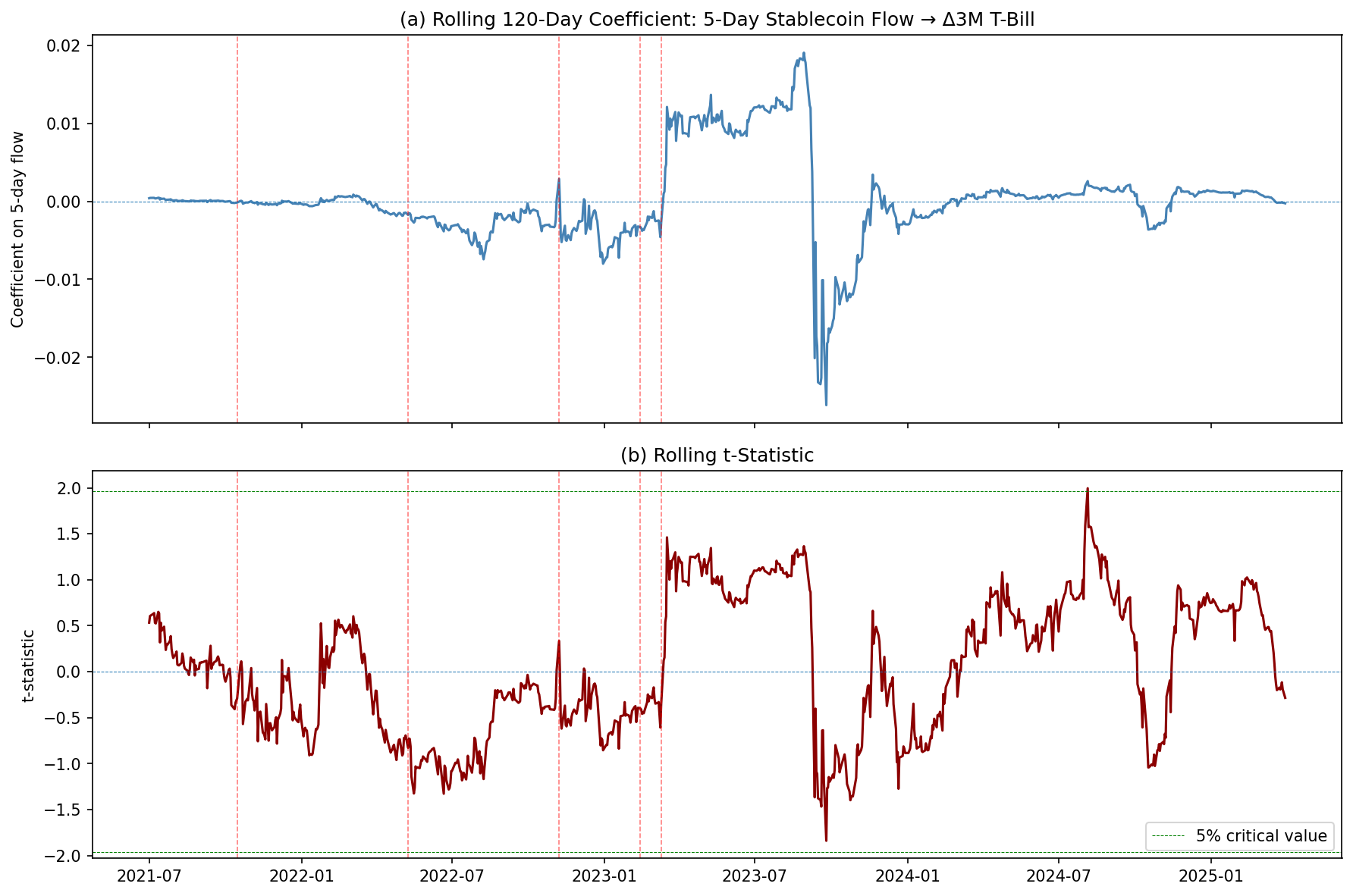}
\caption{Rolling 120-day regression of the change in the three-month bill yield on five-day stablecoin flow. (a) Coefficient. (b) $t$-statistic, with $\pm 1.96$ marked. Vertical lines mark the five episodes.}
\label{c3:fig:rolling}
\end{figure}

The rolling regressions show that there is something to explain. Figure~\ref{c3:fig:rolling} plots the 120-day coefficient on five-day flow and its $t$-statistic. The coefficient sits near zero through 2021 and becomes unstable around the 2022 to 2023 episodes; binary segmentation places breaks on 16~March 2023, within days of the SVB failure and the USDC de-peg, and on 8~September and 6~November 2023. Peg-deviation breaks tell the same story, with USDC breaking around SVB and BUSD breaking later in its wind-down. The relationship is not constant, and it moves when the episodes occur.

\begin{figure}[H]
\centering
\includegraphics[width=\textwidth]{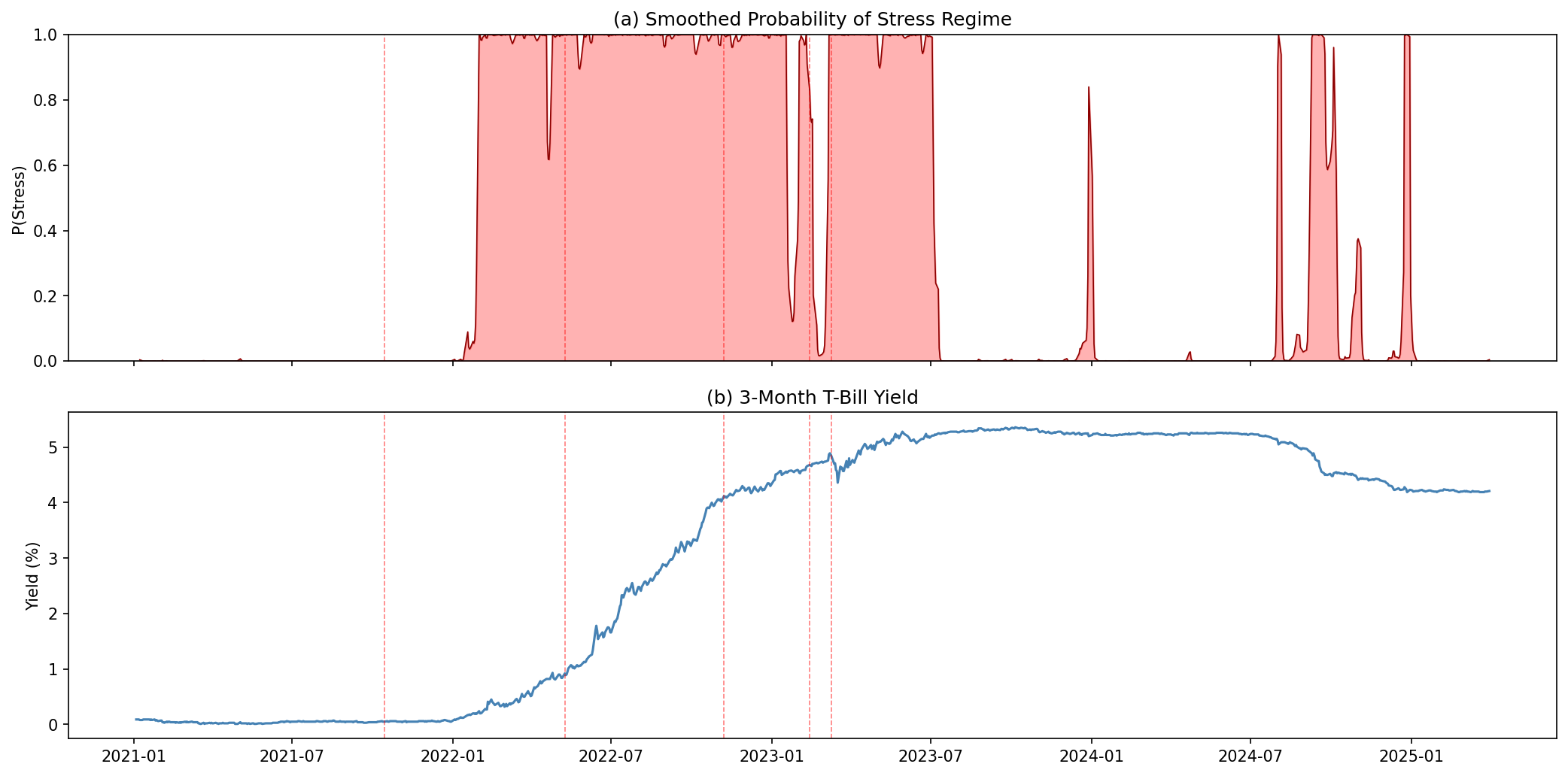}
\caption{Markov-switching model. (a) Smoothed probability of the stress regime. (b) Three-month bill yield. Vertical lines mark the five episodes.}
\label{c3:fig:markov}
\end{figure}

The Markov-switching model shows that the variation is a switch between states rather than drift. Table~\ref{c3:tab:markov} reports two regimes with residual variances of 0.0038 and 0.00012, a ratio of about 32, and stay probabilities of 0.974 and 0.987, which imply expected durations of about 39 days in stress and 78 days in calm. Figure~\ref{c3:fig:markov} shows the smoothed stress probability switching on in early 2022 and staying on through Terra, FTX, the BUSD enforcement and SVB, before switching off in mid-2023. It stays off through the CFTC/Tether window in October 2021, so the one episode the regressions below find uninformative is also the one the regime model does not classify as stress. The regime also switches on briefly in late 2024, around the start of the easing cycle, when no stablecoin event occurred: it is defined by Treasury-yield volatility, not by stablecoin activity, which is why it serves below as an indicator of the state of the Treasury market rather than as a detector of stablecoin stress. The flow coefficient inside the stress regime is not significant in the signed-flow specification, which is expected: the model has one coefficient per regime and no horizon structure, so it can say that a stress state exists but not what happens to yields over the days after a redemption. An outflow-only variant does show a negative and significant coefficient in the stress regime ($-0.0057$ per standard deviation, $p = 0.016$) and none in the calm regime, which anticipates the sign found below. The dynamic response is the job of the local projections.

\begin{table}[H]
\centering
\small
\caption{Markov-switching regression, five-day flow specification.}
\label{c3:tab:markov}
\begin{tabular}{lcccc}
\toprule
& \multicolumn{2}{c}{\textbf{Stress regime}} & \multicolumn{2}{c}{\textbf{Normal regime}} \\
\cmidrule(lr){2-3}\cmidrule(lr){4-5}
& Coef. & $p$-value & Coef. & $p$-value \\
\midrule
Constant & 0.0118 & 0.004 & $-$0.0003 & 0.456 \\
5-day flow ($z$) & $-$0.0008 & 0.867 & $-$0.0002 & 0.671 \\
$\Delta$VIX ($z$) & 0.0080 & 0.082 & 0.0010 & 0.165 \\
S\&P 500 ret ($z$) & 0.0025 & 0.489 & 0.0003 & 0.729 \\
$\Delta$FFR ($z$) & $-$0.0040 & 0.046 & 0.0013 & 0.352 \\
$\sigma^2$ & 0.0038 & $<$0.001 & 0.00012 & $<$0.001 \\
\midrule
Stay probability & \multicolumn{2}{c}{0.974} & \multicolumn{2}{c}{0.987} \\
Expected duration (days) & \multicolumn{2}{c}{39.1} & \multicolumn{2}{c}{78.2} \\
\midrule
Log-likelihood & \multicolumn{4}{c}{2565.1} \\
AIC & \multicolumn{4}{c}{$-$5102.2} \\
$N$ & \multicolumn{4}{c}{1{,}052} \\
\bottomrule
\end{tabular}
\end{table}

Before the state-dependent estimates, the average relationship is worth one sentence: full-sample regressions of the daily bill-yield change on daily flow, five-day flow, outflows alone, and issuance and redemption separately all return insignificant coefficients ($p$ between 0.15 and 0.38), so there is no relationship to find on the average day. Table~\ref{c3:tab:redemption_lp} gives the headline local projections, pooled over all five episodes. The run-period coefficient on a one-billion-dollar five-day redemption is negative through $h=15$ but not significant on its own at any horizon (at $h=10$, $\beta_{\text{run}} = -0.034$, $p = 0.357$), while the run interaction is significant at $h=5$ and $h=10$ ($p = 0.006$ and $0.024$). Read alone, this says that the flow-yield relationship differs between run and normal periods but that the run-period effect is imprecise. The reason is the composition of the run dummy, and the next table shows it.

\begin{table}[H]
\centering
\small
\caption{Local projections, pooled over five episodes: five-day redemption magnitude and the three-month bill yield.}
\label{c3:tab:redemption_lp}
\begin{tabular}{lrrrrrr}
\toprule
\textbf{Horizon} & $\beta_{\text{run}}$ & SE & $t$ & $p_{\text{run}}$ & $\delta$ & $p(\delta)$ \\
\midrule
$h=5$  & $-$0.0348 & 0.0236 & $-$1.475 & 0.140 & $-$0.0787 & 0.006 \\
$h=10$ & $-$0.0337 & 0.0366 & $-$0.921 & 0.357 & $-$0.1291 & 0.024 \\
$h=15$ &  0.0002 & 0.0651 &  0.004 & 0.997 & $-$0.1113 & 0.209 \\
$h=20$ &  0.0662 & 0.0600 &  1.104 & 0.269 & $-$0.0745 & 0.405 \\
\bottomrule
\end{tabular}

\smallskip
\noindent\footnotesize{Dependent variable: cumulative change in the three-month bill yield from $t-1$ to $t+h$. $\beta_{\text{run}}$ is the run-period effect of a one-billion-dollar five-day redemption; $\delta$ is the run interaction. Newey--West standard errors, bandwidth $h+1$.}
\end{table}

\begin{figure}[H]
\centering
\includegraphics[width=\textwidth]{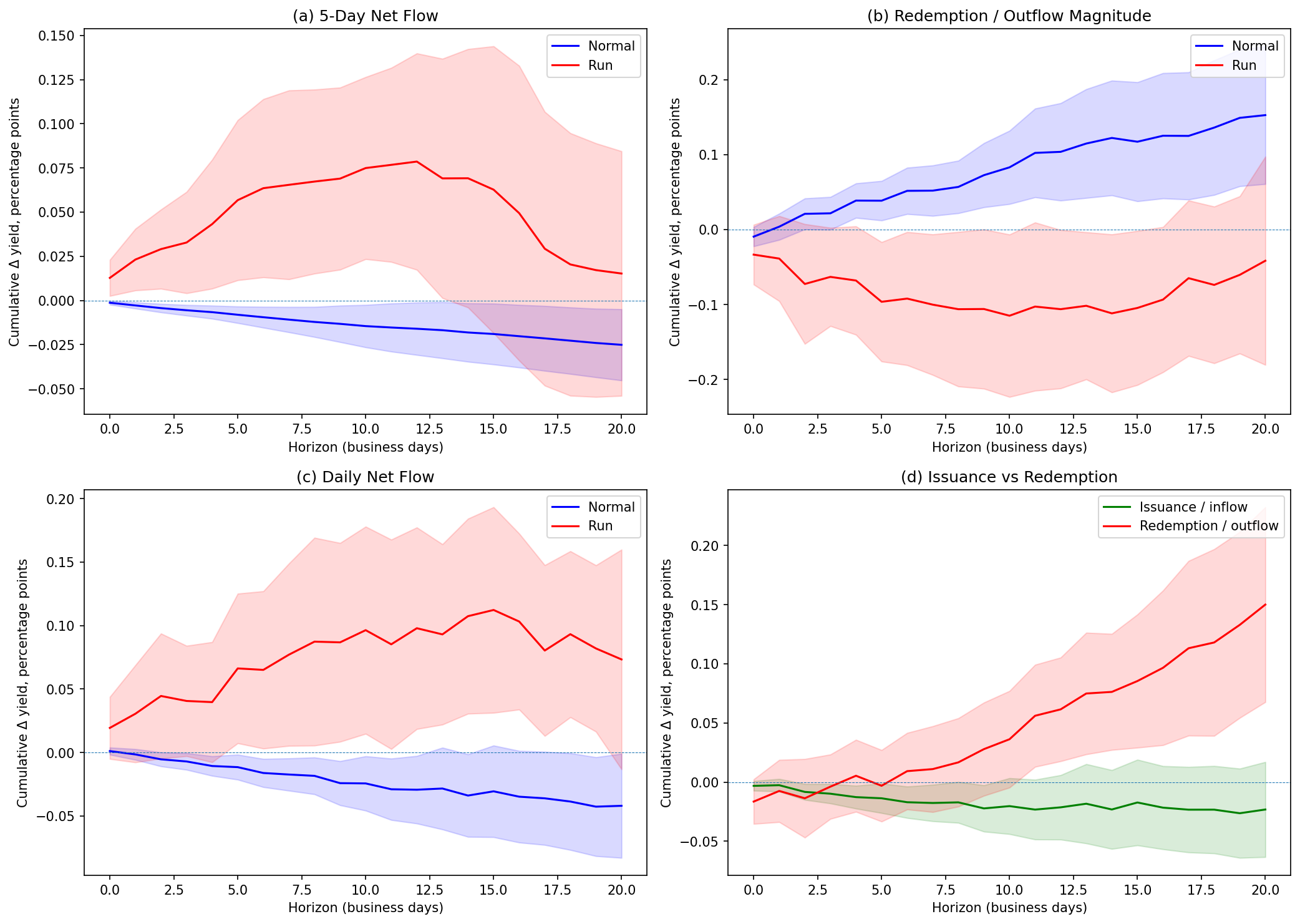}
\caption{Local projection impulse responses, five-day redemption magnitude. Shading: 90\% Newey--West confidence bands.}
\label{c3:fig:lp}
\end{figure}

Table~\ref{c3:tab:panic_vs_reg} splits the run dummy by episode type. In the three panic episodes the effect is clear: at $h=10$ a one-billion-dollar five-day redemption is associated with an 8.5 basis point fall in the three-month bill yield ($\beta_{\text{run}} = -0.085$, $p = 0.026$), with a run interaction significant at the 1\% level ($p = 0.003$), and the same holds at $h=5$. In the two regulatory episodes there is no effect at any horizon: the point estimates at short horizons are positive, none is significant, and the interaction is never significant. Pooling the two types therefore averages an effect of $-0.085$ with an effect of roughly zero, which is why the pooled coefficient in Table~\ref{c3:tab:redemption_lp} falls to $-0.034$ and loses significance while the interaction survives. This is the paper's central empirical result: the signal in stablecoin redemptions comes from panic, not from the size of the redemption as such. A daily redemption specification gives the same sign pattern (full-sample row of Table~\ref{c3:tab:loo}: $\beta_{\text{run}} = -0.082$, interaction $p = 0.010$); the five-day measure remains the headline because it captures sustained pressure. Figure~\ref{c3:fig:lp} shows the impulse responses.

\begin{table}[H]
\centering
\small
\caption{Panic versus regulatory episodes: five-day redemption magnitude.}
\label{c3:tab:panic_vs_reg}
\begin{tabular}{lrrrr}
\toprule
\textbf{Horizon} & \multicolumn{2}{c}{\textbf{Panic (Terra, FTX, SVB)}} & \multicolumn{2}{c}{\textbf{Regulatory (BUSD, CFTC/Tether)}} \\
\cmidrule(lr){2-3}\cmidrule(lr){4-5}
 & $\beta_{\text{run}}$ ($p_{\text{run}}$) & $\delta$ ($p(\delta)$) & $\beta_{\text{run}}$ ($p_{\text{run}}$) & $\delta$ ($p(\delta)$) \\
\midrule
$h=5$  & $-$0.0704 (0.019) & $-$0.1076 (0.001) & 0.0201 (0.346) & 0.0006 (0.979) \\
$h=10$ & $-$0.0853 (0.026) & $-$0.1553 (0.003) & 0.0413 (0.192) & $-$0.0226 (0.574) \\
$h=15$ & $-$0.0181 (0.768) & $-$0.0728 (0.397) & 0.0174 (0.855) & $-$0.1003 (0.333) \\
$h=20$ &  0.0534 (0.308) & $-$0.0222 (0.797) & 0.0749 (0.452) & $-$0.1002 (0.344) \\
\bottomrule
\end{tabular}

\smallskip
\noindent\footnotesize{The headline specification with the run dummy restricted to the named episodes. $p$-values in parentheses. Newey--West standard errors, bandwidth $h+1$.}
\end{table}

The sign decides between the channels. Reserve liquidation predicts that redemptions raise bill yields. In the panic episodes they lower them. Panic-driven runs are embedded in risk-off states in which the demand for safe assets rises by more than any reserve sales could offset, and the maturity evidence in Table~\ref{c3:tab:maturity} says the same thing from a different direction. At $h=10$, pooled over the five episodes, the run-period coefficient is negative at every maturity but weakest where reserves are held: $-3.4$ basis points at three months ($p = 0.357$) and $-5.8$ at six months ($p = 0.177$), against $-14.4$ at two years ($p = 0.001$) and $-10.0$ at ten years ($p = 0.012$). The spread tests in Table~\ref{c3:tab:spreads} confirm it. The six-month minus three-month spread does not move, the ten-year minus three-month spread is only marginal, and the two-year minus three-month spread falls significantly ($p = 0.005$): the two-year yield drops by more than the bill yield during runs, which is the reverse of what selling bills would do. Whatever reserve sales occur, they are not what moves the curve. The response is a broad flight to quality, weighted toward intermediate maturities, in which stablecoin redemptions are one visible symptom.

\begin{table}[H]
\centering
\small
\caption{Maturity segmentation: five-day redemption magnitude at $h=10$, pooled.}
\label{c3:tab:maturity}
\begin{tabular}{lrrrrr}
\toprule
\textbf{Maturity} & $\beta_{\text{run}}$ & SE & $p_{\text{run}}$ & $\delta$ & $p(\delta)$ \\
\midrule
3-month T-bill & $-$0.0337 & 0.0366 & 0.357 & $-$0.1291 & 0.024 \\
6-month T-bill & $-$0.0581 & 0.0431 & 0.177 & $-$0.1214 & 0.035 \\
2-year Treasury & $-$0.1440 & 0.0447 & 0.001 & $-$0.2223 & 0.001 \\
10-year Treasury & $-$0.0998 & 0.0395 & 0.012 & $-$0.1458 & 0.011 \\
\bottomrule
\end{tabular}
\end{table}

\begin{table}[H]
\centering
\small
\caption{Yield-spread tests: five-day redemption magnitude at $h=10$, pooled.}
\label{c3:tab:spreads}
\begin{tabular}{lrrrrr}
\toprule
\textbf{Spread} & $\beta_{\text{run}}$ & SE & $p_{\text{run}}$ & $\delta$ & $p(\delta)$ \\
\midrule
6M minus 3M & $-$0.0245 & 0.0201 & 0.223 & 0.0077 & 0.763 \\
2Y minus 3M & $-$0.1103 & 0.0394 & 0.005 & $-$0.0932 & 0.105 \\
10Y minus 3M & $-$0.0661 & 0.0390 & 0.090 & $-$0.0167 & 0.766 \\
\bottomrule
\end{tabular}
\end{table}

How much of this rests on one episode? Table~\ref{c3:tab:episodes} gives the episode-by-episode projections. Terra is negative at every horizon, with the interaction significant at $h=10$ and $h=15$. SVB is negative and significant at $h=5$ and $h=10$; BUSD enforcement is negative and significant at $h=5$ and marginal at $h=15$; FTX is small and significant only at $h=15$. CFTC/Tether contributes nothing: its coefficients are within a rounding error of zero, and the significant interaction $p$-values at longer horizons should not be read as evidence of an effect, since the episode contains almost no redemption days and the interaction is estimated on next to no variation. Averaging the $h=10$ coefficients, the three panic episodes give $-0.113$ with two of three individually significant, and the two regulatory episodes give $-0.015$ with neither significant, which is the same split as Table~\ref{c3:tab:panic_vs_reg} from a different angle. The leave-one-out check in Table~\ref{c3:tab:loo} states the dependence on Terra plainly: dropping Terra takes the run coefficient to $0.005$ ($p = 0.866$) and the interaction to $-0.045$ ($p = 0.271$); dropping any other episode leaves both negative and the interaction significant. The result is therefore a Terra-led finding corroborated by SVB and, more weakly, by BUSD and FTX, and none of the smaller episodes would establish it alone. That is how it is read throughout.

\begin{table}[H]
\centering
\small
\caption{Episode-by-episode local projections: redemption magnitude.}
\label{c3:tab:episodes}
\begin{tabular}{lrrrrrr}
\toprule
& \multicolumn{2}{c}{$h=5$} & \multicolumn{2}{c}{$h=10$} & \multicolumn{2}{c}{$h=15$} \\
\cmidrule(lr){2-3}\cmidrule(lr){4-5}\cmidrule(lr){6-7}
\textbf{Episode} & $\beta_{\text{ep}}$ & $p(\delta)$ & $\beta_{\text{ep}}$ & $p(\delta)$ & $\beta_{\text{ep}}$ & $p(\delta)$ \\
\midrule
Terra & $-$0.1427 & 0.063$^{*}$ & $-$0.2261 & 0.019$^{**}$ & $-$0.2105 & 0.003$^{***}$ \\
FTX & 0.0130 & 0.759 & $-$0.0193 & 0.184 & $-$0.0355 & 0.017$^{**}$ \\
BUSD enforcement & $-$0.1209 & 0.012$^{**}$ & $-$0.0300 & 0.438 & $-$0.0734 & 0.064$^{*}$ \\
SVB & $-$0.1692 & 0.025$^{**}$ & $-$0.0940 & 0.036$^{**}$ & 0.0032 & 0.232 \\
CFTC/Tether & 0.0070 & 0.889 & 0.0001 & 0.097$^{*}$ & $-$0.0001 & 0.013$^{**}$ \\
\bottomrule
\end{tabular}

\smallskip
\noindent\footnotesize{$\beta_{\text{ep}}$ is the episode-specific run-period redemption effect; $p(\delta)$ is the $p$-value on the episode interaction. $^{*}$~$p<0.10$, $^{**}$~$p<0.05$, $^{***}$~$p<0.01$.}
\end{table}

\begin{table}[H]
\centering
\small
\caption{Leave-one-out robustness: daily redemption magnitude at $h=10$.}
\label{c3:tab:loo}
\begin{tabular}{lrrrrr}
\toprule
\textbf{Dropped episode} & $\beta_{\text{run}}$ & $p_{\text{run}}$ & $\delta$ & $p(\delta)$ & $N$ \\
\midrule
None (full sample) & $-$0.0816 & 0.139 & $-$0.1707 & 0.0098$^{***}$ & 1{,}040 \\
Drop Terra & 0.0053 & 0.866 & $-$0.0451 & 0.271 & 1{,}040 \\
Drop FTX & $-$0.0924 & 0.153 & $-$0.1864 & 0.012$^{**}$ & 1{,}040 \\
Drop BUSD enforcement & $-$0.0959 & 0.102 & $-$0.1799 & 0.008$^{***}$ & 1{,}040 \\
Drop SVB & $-$0.0913 & 0.128 & $-$0.1775 & 0.012$^{**}$ & 1{,}040 \\
Drop CFTC/Tether & $-$0.1150 & 0.081 & $-$0.1980 & 0.008$^{***}$ & 1{,}040 \\
\bottomrule
\end{tabular}
\end{table}

The window checks show that the effect is a sustained-stress phenomenon, not an announcement-day reaction. Table~\ref{c3:tab:event_windows} uses forward-looking windows of 5 to 42 business days after each event start. Pooled over five episodes, $\beta_{\text{run}}$ is never significant on its own, though the interaction is at 20 and 42 days. Restricted to the panic episodes, the pattern the baseline window was chosen to capture appears directly: nothing at 5 or 10 days, then $\beta_{\text{run}} = -0.105$ ($p = 0.019$) at 20 days and $-0.083$ ($p = 0.031$) at 42 days, with interactions significant at the 1\% level. The relationship builds over weeks as the stress regime takes hold. The same exercise on the regulatory episodes alone gives positive coefficients in the 5 and 10 day windows, on 10 and 20 run observations with almost no redemptions, and nothing at 20 or 42 days; the wrong sign on a handful of observations is not evidence of an effect, but it confirms that nothing in the regulatory episodes pushes in the panic direction.\footnote{The same forward-window exercise on the two regulatory episodes gives small positive coefficients at 5 and 10 days with implausibly small standard errors, because those windows contain almost no redemption days and the coefficient is estimated on next to no variation; at 20 and 42 days nothing is significant. None of it indicates an effect.}

\begin{table}[H]
\centering
\small
\caption{Event-window robustness: five-day redemption magnitude at $h=10$.}
\label{c3:tab:event_windows}
\begin{tabular}{llrrrrrr}
\toprule
\textbf{Sample} & \textbf{Window} & \textbf{Run obs.} & $\beta_{\text{run}}$ & SE & $p_{\text{run}}$ & $\delta$ & $p(\delta)$ \\
\midrule
Pooled (5 episodes) & 5 days  & 25  & $-$0.0091 & 0.0275 & 0.741 & $-$0.0722 & 0.030 \\
Pooled (5 episodes) & 10 days & 50  & 0.0290 & 0.0303 & 0.338 & $-$0.0365 & 0.377 \\
Pooled (5 episodes) & 20 days & 98  & $-$0.0393 & 0.0419 & 0.349 & $-$0.1362 & 0.005 \\
Pooled (5 episodes) & 42 days & 186 & $-$0.0197 & 0.0344 & 0.568 & $-$0.1147 & 0.040 \\
\addlinespace
Panic only & 5 days  & 15  & $-$0.0287 & 0.0205 & 0.162 & $-$0.0910 & 0.002 \\
Panic only & 10 days & 30  & $-$0.0015 & 0.0324 & 0.963 & $-$0.0638 & 0.130 \\
Panic only & 20 days & 60  & $-$0.1045 & 0.0445 & 0.019 & $-$0.1824 & $<$0.001 \\
Panic only & 42 days & 126 & $-$0.0827 & 0.0383 & 0.031 & $-$0.1533 & 0.003 \\
\bottomrule
\end{tabular}

\smallskip
\noindent\footnotesize{Forward-looking windows from each event start date. Newey--West standard errors, bandwidth $h+1$.}
\end{table}

Table~\ref{c3:tab:placebo} compares those estimates with matched placebo windows. Pooled over five episodes, the 42-day estimate is not distinguishable from random windows (two-sided $p = 0.588$, lower-tail $p = 0.069$). Restricted to the panic episodes, the 42-day coefficient is unusually negative relative to randomly placed windows with the same number of run observations: the lower-tail placebo $p$-value is 0.015, and the two-sided value of 0.079 is marginal. The placebo test does not establish causality, but it shows that the panic-window estimate is not a generic feature of arbitrary stretches of the sample.

\begin{table}[H]
\centering
\small
\caption{Matched placebo-window robustness.}
\label{c3:tab:placebo}
\setlength{\tabcolsep}{4pt}
\resizebox{\textwidth}{!}{%
\begin{tabular}{llrrrrrrr}
\toprule
\textbf{Sample} & \textbf{Window} & \textbf{Run obs.} & $\beta_{\text{run}}$ & $p_{\text{run}}$ & $\delta$ & $p(\delta)$ & \textbf{Two-sided $p$} & \textbf{Lower-tail $p$} \\
\midrule
Pooled (5 episodes) & 20 days & 98  & $-$0.0393 & 0.349 & $-$0.1362 & 0.005 & 0.572 & 0.102 \\
Pooled (5 episodes) & 42 days & 186 & $-$0.0197 & 0.568 & $-$0.1147 & 0.040 & 0.588 & 0.069 \\
Panic only & 20 days & 60  & $-$0.1045 & 0.019 & $-$0.1824 & $<$0.001 & 0.298 & 0.055 \\
Panic only & 42 days & 126 & $-$0.0827 & 0.031 & $-$0.1533 & 0.003 & 0.079 & 0.015 \\
\bottomrule
\end{tabular}}

\smallskip
\noindent\footnotesize{Placebo $p$-values compare the actual run coefficient with its distribution across randomly assigned non-event windows matched on the number of run observations.}
\end{table}

Two further checks address confounding. Table~\ref{c3:tab:calendar_excl} drops FOMC decision days and CPI release days, the acute SVB banking-stress window of 8 to 24~March 2023, and both. The run coefficient grows in magnitude under every exclusion, from $-0.034$ to $-0.058$ with both applied, and the interaction stays significant throughout. Removing the SVB bank panic, the case most likely to show that the SVB contribution was a banking artefact rather than a stablecoin signal, strengthens the estimate. The result is therefore not driven by announcement days or by the acute banking-stress window.

\begin{table}[H]
\centering
\small
\caption{Calendar-exclusion robustness: five-day redemption magnitude at $h=10$, pooled.}
\label{c3:tab:calendar_excl}
\begin{tabular}{lrrrrr}
\toprule
\textbf{Sample} & $\beta_{\text{run}}$ & $p_{\text{run}}$ & $\delta$ & $p(\delta)$ & $N$ \\
\midrule
Baseline                            & $-$0.0337 & 0.357 & $-$0.1291 & 0.024 & 1{,}040 \\
Exclude FOMC and CPI days           & $-$0.0434 & 0.272 & $-$0.1484 & 0.019 & 956 \\
Exclude SVB banking-stress window   & $-$0.0508 & 0.196 & $-$0.1406 & 0.016 & 1{,}027 \\
Exclude both                        & $-$0.0580 & 0.171 & $-$0.1566 & 0.015 & 945 \\
\bottomrule
\end{tabular}

\smallskip
\noindent\footnotesize{The SVB banking-stress window (8--24~March 2023) is narrower than the USDC/SVB run window of Table~\ref{c3:tab:events}. Newey--West standard errors, bandwidth $h+1$.}
\end{table}

The issuer-level comparison in Table~\ref{c3:tab:svb} adds a last piece. Around SVB, the daily USDC flow interaction is significant and the USDT placebo is not, and in a horse race USDC carries the effect while USDT does not; the five-day USDC interaction, which matches the paper's main horizon, is not significant. The comparison shows issuer-specific stress and substitution within the sector, but the same weeks contain a bank run and large Treasury moves, so it is descriptive support and not a causal estimate.

\begin{table}[H]
\centering
\small
\caption{USDC/SVB issuer-level comparison.}
\label{c3:tab:svb}
\begin{tabular}{lrrrr}
\toprule
\textbf{Specification} & \textbf{Interaction} & \textbf{SE} & $p$ & $N$ \\
\midrule
USDC 5-day flow $\times$ Post & 0.008 & 0.016 & 0.621 & 122 \\
USDC daily flow $\times$ Post & 0.139 & 0.038 & $<$0.001$^{***}$ & 119 \\
USDT daily flow $\times$ Post (placebo) & $-$0.069 & 0.043 & 0.107 & 119 \\
Horse race: USDC $\times$ Post & 0.153 & 0.045 & $<$0.001$^{***}$ & 119 \\
Horse race: USDT $\times$ Post & 0.016 & 0.041 & 0.692 & 119 \\
\bottomrule
\end{tabular}
\end{table}

Taken together, the results support one reading and rule out another. What is ruled out is reserve liquidation as the dominant force: the sign is wrong for it, the response is broad rather than confined to bills, and the volumes involved are small against the Treasury market. What is supported is a surveillance reading. Stablecoin redemptions are informative about Treasury-market conditions when, and only when, they are driven by panic; the information builds over weeks rather than days; and it takes the form of a broad flight to quality in which the stablecoin sector is a fast, visible participant. The evidence has clear limits, and they are stated here rather than deferred. There are three panic episodes, and the estimate depends on the largest of them. Runs and yield compression may share a common cause that no finite set of controls removes, so the paper shows correlation within stress regimes and does not claim causation. And the design covers three dollar tokens and one sovereign market. What the paper can say is that a central bank watching stablecoin redemptions during a panic is watching something informative about the safe-asset market, and that a central bank watching them during a regulatory wind-down is not.

\section{Policy Recommendations}\label{c3:sec:policy}

Four recommendations follow.

\paragraph{Measure stablecoin flows in units, not market capitalisation.} The standard proxy fails exactly when it is needed. Any monitoring system, and any future empirical work, should compute flows from units outstanding at par as in equation~\eqref{c3:eq:corrected_flow}. The correction costs nothing and removes phantom inflows during de-pegs.

\paragraph{Monitor redemptions as a stress signal, not as a yield forecast.} Large redemptions during a panic do not mean bill yields are about to rise; on this evidence they accompany yield compression across the curve. The right use of the flow series is as a high-frequency marker that a safe-asset stress regime has begun. Five-day aggregate redemptions of roughly \$2 to 3 billion across the major tokens are the scale reached in the three panic episodes and the scale at which the signal becomes informative. At the panic-subsample coefficient, a \$3 billion five-day redemption in a panic-driven regime is associated with a cumulative fall of about 26 basis points in the three-month bill yield within ten days, which is large enough to matter for money-market pricing, collateral valuation and the reading of short-rate movements.

\paragraph{Condition the signal on the type of episode.} The same redemption volume carries no information during a regulatory wind-down. A dashboard should pair the flow series with two indicators that separate panic from orderly contraction: the maximum absolute peg deviation across the major tokens, which identifies when a token has become information-sensitive, and a real-time stress-regime probability of the kind the Markov-switching model produces, which summarises whether the market is in a state where flow signals are likely to be informative. Redemptions that occur without peg stress and outside a high-volatility regime, as in the CFTC/Tether and BUSD episodes, should be discounted.

\paragraph{Watch the whole curve.} Because the response is broad and weighted toward intermediate maturities, monitoring only the three-month bill on the assumption that reserves sit there would miss most of the movement. Two-year and ten-year yields, and the two-year minus three-month spread, belong in the same dashboard.

Two cautions apply. The thresholds are calibrated on three panic episodes dominated by Terra, so they should be re-estimated as new episodes accumulate rather than fixed. And the signal is informational: it tells a supervisor that a stress state is under way, not what caused it or what will follow.

\section{Conclusion}\label{c3:sec:conclusion}

This paper asked whether stablecoin run episodes carry information about U.S.\ Treasury yields, and found that they do, on two conditions: the episode must be panic-driven, and the information must be read as a stress signal rather than as a mechanical channel.

Three results support that answer. First, the relationship is state-dependent. Stablecoin flows have no stable relationship with yields in normal periods, the data contain a stress regime whose residual variance is thirty-two times the calm regime's, and within the three panic-driven runs a one-billion-dollar five-day redemption is associated with an 8.5 basis point fall in the three-month bill yield within ten days, building over twenty to forty-two days and standing out against matched placebo windows. Second, the signal depends on the kind of episode. Two regulatory actions, one of them the wind-down of a top-three token, leave no trace in yields, and pooling them with the panic episodes hides the effect. Third, the mechanism is flight to quality, not reserve liquidation: redemptions accompany lower yields, not higher, and the response is broad across the curve and weighted away from the bills reserves hold. The paper also supplies a measurement correction that future work on stablecoin flows should adopt, since the standard market-capitalisation proxy records phantom inflows in exactly the windows a study of runs depends on.

The result depends heavily on Terra, corroborated by SVB and more weakly by FTX and BUSD, and the paper says so wherever the estimate is used. It shows co-movement within stress regimes rather than causation, and it covers three tokens and one sovereign market. Those limits shape the claim but do not undermine it, because the claim is about surveillance: a central bank that watches units-based stablecoin redemptions, conditions on peg stress and a stress-regime indicator, and reads a \$2 to 3 billion five-day redemption during a panic as a marker that safe-asset demand is intensifying, is using a fast and freely available signal that the results here show to be informative.

The stablecoin sector is now large enough that its runs are no longer contained crypto-market events; they sit at the interface between money-like claims, sovereign safe assets and central bank monitoring. Extending the evidence to future episodes as they occur, using intraday data to separate reserve sales from contemporaneous flight to quality, and feeding a stablecoin stress indicator into macro-financial scenario models used for stress testing are the natural next steps.

\section*{Declarations}

\paragraph{Funding.} Ahmad Qadi was supported by a doctoral scholarship from the Saudi Arabian Ministry of Education. This research did not receive any other specific grant from funding agencies in the public, commercial or not-for-profit sectors.

\paragraph{Competing interests.} The authors declare that they have no known competing financial interests or personal relationships that could have appeared to influence the work reported in this paper.

\paragraph{Data availability.} All data are public: stablecoin prices and market capitalisation from CoinGecko, and Treasury yields and control variables from the Federal Reserve Economic Data (FRED) database. The analysis code is available from the corresponding author on request.

\paragraph{Use of generative AI.} During the preparation of this work, the authors used Claude (Anthropic) and ChatGPT (OpenAI) to assist with parts of the text, check reported numbers against tables and code output, verify bibliographic details and debug LaTeX and code. All formulas, empirical claims, references, notation and manuscript content were reviewed and verified by the authors, who take full responsibility for the content of this manuscript.

\end{document}